\documentclass[prb,twocolumn,superscriptaddress,showpacs]{revtex4}
\usepackage[ansinew]{inputenc}
\usepackage{amsmath}
\usepackage{amsfonts}
\usepackage{graphicx}
\usepackage{hyperref}
\usepackage{dcolumn}
\begin{document}
\newcommand{\phantomFrac}{\vphantom{\frac{}{}}}
\newcommand{\e}{\varepsilon}
\newcommand{\subst}{\left.\phantomFrac\right|}
\newcommand{\imag}{\Im m\;}
\newcommand{\real}{\Re e\;}
\newcommand{\order}{\mathcal{O}}
\newcommand{\Az}[1][n]{A^{(#1)}}
\newcommand{\Jz}[2][n]{J_{#2}^{(#1)}}
\newcommand{\Atr}[1][n]{A_\text{Tr}^{(#1)}}
\newcommand{\An}[1][n]{{A_\text{N}}^{(#1)}}
\newcommand{\Ai}[1][n]{{A_\text{I}}^{(#1)}}
\newcommand{\Ae}[1][n]{\varepsilon_{A}}
\newcommand{\Aa}[1][n]{\chi^{(#1)}}
\newcommand{\Fi}[1][n]{\ensuremath{{f_\text{I}}^{(#1)}}}
\newcommand{\Fn}[1][n]{\ensuremath{{f_\text{N}}^{(#1)}}}

\title{Bloch gain in dc-ac-driven semiconductor superlattices in the
absence of electric domains}
\author{Timo Hyart}
\affiliation{Department of Physical
Sciences, P.O. Box 3000, FI-90014 University of Oulu, Finland}
\author{Kirill N. Alekseev}
\affiliation{Department of Physical Sciences, P.O. Box 3000,
FI-90014 University of Oulu, Finland}
\affiliation{Department of Physics, Loughborough University, LE11
3TU, United Kingdom}

\author{Erkki V. Thuneberg}
\affiliation{Department of Physical
Sciences, P.O. Box 3000, FI-90014 University of Oulu, Finland}
\begin{abstract}
We study theoretically the feasibility of amplification and
generation of terahertz radiation in dc-ac-driven semiconductor
superlattices in the absence of electric domains. We find that if in
addition to dc bias a strong THz pump field is applied, Bloch gain
profile for a small THz signal can be achieved under conditions of
positive static differential conductivity. Here the positive
differential conductivity arises, similarly to the case of
large-signal amplification scheme [H. Kroemer, cond-mat/0009311)],
due to modifications of dc current density caused by the application
of high-frequency ac field [K. Unterrainer \textit{et al.}, Phys.
Rev. Lett. \textbf{76}, 2973 (1996)]. Whereas the sign of absorption
at low and zero frequencies is sensitive to the ac fields, the gain
profile in the vicinity of gain maximum is robust. We suggest to use
this ac-induced effect in a starter for THz Bloch oscillator. Our
analysis demonstrates that the application of a short THz pulse to a
superlattice allows to suppress the undesirable formation of
electric domains and reach a sustained large-amplitude operation of
the dc-biased Bloch oscillator.
\end{abstract}



\maketitle

\section{Introduction}

Terahertz radiation ($0.3-10$ THz) has many promising applications
in different areas of science and technology such as space
astronomy, wideband communications and biosecurity
\cite{thz-reviews}. One of the main challenges in the development of
THz technology is the construction of  coherent, monochromatic and
miniature sources of THz radiation that can operate at room
temperature. A great recent achievement was the development of
quantum cascade lasers that operate at THz frequencies employing
quantum transitions between energy levels in multiple quantum well
heterostructures \cite{nature-QCL}. Continuous improvements in the
design of THz quantum cascade lasers have allowed to increase their
operational temperature above $100$ K \cite{williams}. However, it
is certainly very difficult to maintain population inversion in a
THz quantum cascade laser at room temperature.
\par
Semiconductor superlattices (SLs) \cite{Esaki}, which are working
in the miniband transport regime, attract much attention as an
artificial nonlinear medium demonstrating a resonant interaction
with THz radiation (for review, see
\cite{bass86,wackerrew,platero04}). In the presence of dc bias,
electrons from a single miniband of SL can perform transient THz
Bloch oscillations \cite{Bloch,Zener}, which decay at a picosecond
time scale due to unavoidable scattering \cite{exper_B-osc}. In
the stationary transport regime, when the scattering of the
miniband electrons is an important factor, the possibility of
inversionless amplification and generation of THz radiation in SLs
has been well recognized after the seminal suggestions
\cite{Esaki,KSS}. Bloch oscillations can potentially lead to the
amplification of a weak ac field at frequencies smaller than the
Bloch frequency, whereas absorption occurs for frequencies which
are larger than the Bloch frequency \cite{KSS}. If electrons can
perform relatively many cycles of Bloch oscillations between
scattering events, the maximum of the small-signal gain is
achieved in the vicinity of the Bloch frequency. This Bloch gain
profile, shaped as a familiar dispersion curve, has recently
attracted a lot of attention because estimates predicted a
significant THz gain near the Bloch frequency at room temperature.
Moreover, Bloch gain for THz signals of a large amplitude has also
been estimated \cite{romanov78,dodin93}.
\par
Within the semiclassical picture, the physical mechanism of the
high-frequency gain can be understood in terms of ballistic
trajectories of electrons in the quasimomentum space. It was found
that the miniband electrons are gathered into certain favorable
trajectories forming electron bunches, which are eventually
responsible for the negative absorption \cite{Kroemer2}.
Alternatively, the Bloch gain can be explained within theories which
consider the scattering-assisted transitions between the
Wannier-Stark states \cite{willenberi}. It is likely that the second
order scattering-assisted mechanism of gain is not restricted to the
case of single-band transport in SLs. The dispersive profile of
optical gain  has been recently observed in the quantum cascade
lasers \cite{Faist}.
\par
However, at least in the case of miniband transport in dc-biased
SL, the small-signal gain profile extends to arbitrary low
frequencies \cite{KSS}, which in essence means that the dc
differential conductivity is also negative. SLs with static
negative differential conductivity (NDC), similarly to the bulk
semiconductors demonstrating  Gunn effect \cite{Gunn}, are
unstable against the development of space-charge instability
\cite{KSS,buettiker77-ignatov87}. This electric instability
eventually results in the formation of moving domains of high
electric field inside the SL \cite{french-renk-domains_exp}. So
far, the NDC-related instability has prevented an unambiguous
observation of the net Bloch gain in long SLs.
\par
Importantly, it is still potentially possible to avoid the formation
of the electric domains in SL, if the amplified THz signal has
enough large amplitude. In this case, as has been demonstrated for a
particular set of amplitude and frequency \cite{Kroemer}, the
dependence of the time-averaged current on the applied bias can have
a positive slope for a proper choice of the operation point. This
effect in SLs is similar to the so-called Limited Space-charge
Accumulation (LSA) mode of operation known in semiconductor devices
with hot electrons \cite{copeland,kagan,sokolov}. The finding
\cite{Kroemer} can potentially be used for the construction of a THz
amplifier with a large offset. However, to realize the THz
oscillator it is still necessary to understand how to avoid the
destructive electric domains at small amplitude in order to reach
the large-amplitude regime supporting stable operation of the
device. This is known as the device-turn-on problem for THz Bloch
oscillator \cite{Kroemer}.
\par
In recent years a number of interesting suggestions for the
realization of small-signal Bloch gain in a SL were put forward. The
first direction in this research area consists in the experimental
investigations of the Bloch gain at short space or time scales, when
the electric domains have not enough time to build up. Advanced
nanostructure design, an array of short SLs, has been introduced to
measure a frequency dependent crossover from loss to gain
\cite{savvidis}. Modern ultrafast optical techniques were applied to
measure the Bloch gain in undoped superlattices during a short time
window after a femtosecond optical excitation of carriers in SL
\cite{lisauskas,Hirakawa}. Another interesting suggestion is to work
with 2D structures, where the electric domains are effectively more
suppressed in comparison with the case of 3D structures
\cite{2D_domains,Feil_another}. With this aim a lateral surface SL
shunted by another SL \cite{sl_shunt} have been introduced as an
active medium for a future realization of the Bloch oscillator
\cite{Feil_main}. At least theoretically, it is also possible to
obtain a high-frequency gain in conditions of a positive dc
differential conductivity (PDC) by engineering the miniband
dispersion relations in SLs \cite{Romanovdispersion} or using a hot
electron injection into a miniband \cite{hotel}.
\par
In this paper, we show theoretically how to reach gain for THz
signals of both small and large amplitudes in a single miniband SL
combining the action of constant and alternating electric fields. We
focus on the requirements for gain under conditions of the electric
stability. With this aim we first suppose that the electric field,
acting on the miniband electrons, is $E_{dc}+E_{\delta}\cos{\omega
t}$ and re-examine the scheme of electrically stable
\textit{large-amplitude} ($E_{\delta}$) operation of Bloch
oscillator \cite{Kroemer}. We demonstrate that the gain in
conditions of static PDC exists in rather wide ranges of $\omega$
and $E_{\delta}$. In a real device, the ac probe field
$E_{\delta}\cos{\omega t}$ should be the mode of a high-$Q$
resonator.
\par
As a next step, we consider the action of the bichromatic field
$E_{dc}+E_{ac}\cos{\Omega t}+E_\delta \cos{\omega t}$, where
$E_{ac}\cos{\Omega t}$ is an additional external THz field (pump)
and the amplitude of the probe field (resonator mode), $E_{\delta}$,
\textit{is now small}. The pump $\Omega$ and probe $\omega$
frequencies are assumed to be incommensurate. The basic idea behind
this part of our paper is to use the external ac field to suppress
the electric instability in SL for some range of bias $E_{dc}$. We
show that in the case of dc-ac-driven SL the dispersive gain profile
stays almost unaltered at high-frequencies $\omega\simeq n\Omega$
($n=1,2,\ldots$), although at small frequencies the absorption can
change its sign and become positive (see Fig.~\ref{spabsvsx2}).
Thus, THz gain without the electric instability can be achieved, and
this bichromatic scheme can potentially be used to realize a THz
generator. The expense we must pay for this new attractive
possibility is the necessity to use the pump THz field.
\par
We find that the physical mechanism of suppression of the electric
instability in both these schemes is based on local modifications of
the voltage-current (VI) characteristics of the SL induced by the
THz field \cite{theory_old,unterrainer}. However, while in the
large-amplitude mode of the Bloch oscillator such changes in the dc
current are self-induced by the strong resonator field
$E_{\delta}\cos{\omega t}$, in the bichromatic scheme, at least in
the limit of weak probe field ($E_{\delta}\rightarrow 0$), they are
mainly caused by the action of the external ac field.
\par
Our analysis also shows that it is potentially possible to combine
these two schemes of generation in a single device. Namely, we
suggest to use the external field as \textit{a starter for THz Bloch
oscillator}: Temporal action of the ac pump can provide both the
necessary suppression of electric domains and the gain for the
small-amplitude oscillations of the resonator mode, before the field
strength in this mode, $E_{\delta}$, can reach up to the amplitude
supporting electrically stable operation without external pump.
Since the incommensurability of frequencies is supposed, the pump
field of the starter should not necessarily be monochromatic.
Intensive, broadband THz pulses (T-rays) \cite{T-rays-rev} can be
also used to ignite the stable large-amplitude operation of the
superlattice Bloch oscillator. Thus our theoretical research
contributes to the solution of the device-turn-on problem of the
canonic Bloch oscillator.

\section{\label{sec1} Nonlinear electron transport and the canonic Bloch oscillator}

Everywhere in this paper we work within the semiclassical approach
based on the use of Boltzmann transport equation for the miniband
with the tight-binding dispersion relation \cite{bass86,wackerrew}.
We mainly use the standard approximation of a single constant
relaxation time $\tau$ and employ an exact formal solution of the
Boltzmann equation \cite{chambers}; the effects of two distinct
relaxation times are discussed in Appendix. We are interested in the
electron dynamics under the action of a time-dependent field which
consists of a constant and alternating parts
$E(t)=E_{dc}+E_{alt}(t)$. In general case, the ac part $E_{alt}(t)$
can contain many incommensurate frequency components. We define the
absorption of the probe ac field $E_\delta \cos\omega t$ in the SL
miniband as
\begin{equation}
\label{absorb-def} \mathcal{A}(\omega)=\langle j(t)\cos{\omega
t}\rangle_t,
\end{equation}
where $j(t)$ is the current density induced in the SL by the total
field $E(t)$. In general case the averaging $\langle\ldots\rangle_t$
is performed over infinite time. Gain corresponds to
$\mathcal{A}<0$. In this section we will consider only the
monochromatic ac field
\begin{equation}
\label{E-mono}
E(t)=E_{dc}+E_{\delta} \cos{\omega t}.
\end{equation}
For this field averaging in Eq.~(\ref{absorb-def}) should be
performed only over the period $T=2\pi/\omega$.
\par
The absorption (gain) $\alpha$ in units $\rm{cm}^{-1}$ is related
to $\mathcal{A}$ as \cite{willenberi}
$$
\alpha=\frac{2}{n_r \epsilon_0 c} \frac{\mathcal{A}}{E_{\delta}},
$$
where $\epsilon_0$ and $c$ are the permittivity and the speed of
light in vacuum, and $n_r$ is the refractive index of SL material.
Whenever the gain overcomes the loss in the resonator, we have an
oscillator. For the oscillator the generated power density
$\mathcal{P}$ inside the sample is
$$
\mathcal{P}=\mathcal{A} E_\delta .
$$
To estimate the gain and generated power in physical units we will
use everywhere in this paper the following parameters of typical
GaAs/AlAs SL: period $d=6 \ \rm{nm}$, miniband width $\Delta= 60 \
\rm{meV}$, density of electrons $N=10^{16} \ \rm{cm}^{-3}$,
refractive index $n_r=\sqrt{13}$ (GaAs). We consider operation at
room temperature. The characteristic scattering time of miniband
electrons is $\tau\simeq 200$ fs.
\par
Before proceeding with the analysis of gain in ac-driven case, it is
worth to remind the main nonlinear transport properties of a
dc-biased SL. The dependence of the dc current density on the dc
bias is given by the Esaki-Tsu formula \cite{Esaki}
\begin{equation}
\label{eq:ET} j_{dc}(e E_{dc} d)=j_{\rm{peak}} \frac{2 e E_{dc}
d/\Gamma}{1+(e E_{dc} d/\Gamma)^2}.
\end{equation}
Here $\Gamma=\hbar/\tau$  and the peak current, corresponding to
$E_{dc}=E_{cr}$ ($E_{cr}\equiv\hbar/(ed\tau)$), is
$$
j_{\rm{peak}}=\frac{e N v_0}{2}
\frac{I_1(\frac{\Delta}{2k_bT})}{I_0(\frac{\Delta}{2k_bT})},
$$
where $v_0=\Delta d/(2 \hbar)$ is the maximal electron velocity in
the miniband and $I_k(x)$ ($k=0,1$) are the modified Bessel
functions. This temperature dependence of dc current is in a good
agreement with the experiment \cite{allen90}. The dependence of
$j_{dc}$ on $E_{dc}$ is shown in Fig.~\ref{fig:KSSkuva}. Instead
of the field strength variable $E_{dc}$, the Esaki-Tsu
characteristic can be represented in terms of Bloch frequency
$\omega_B=edE_{dc}/\hbar$ using the equality
$E_{dc}/E_{cr}=\omega_B \tau$. For $\omega_B \tau \sim 1$ the
Bloch frequency belongs to THz frequency range \cite{Esaki}.
\par
The dc differential conductivity, $\sigma_{dc}=\partial
j_{dc}/\partial E_{dc}$, defines the slope of VI characteristic at
the working point. For the Esaki-Tsu dependence (\ref{eq:ET}) it is
\begin{equation}
\sigma_{dc}(eE_{dc}d)=\frac{2 j_{\rm peak}}{E_{cr}} \frac{1-(e
E_{dc}d)^2/\Gamma^2}{\big[1+(e E_{dc} d)^2/\Gamma^2 \big]^2}.
\label{esakidiffcond}
\end{equation}
If $E_{dc}/E_{cr}=\omega_B \tau >1$, $\sigma_{dc}$ is negative and
therefore Esaki-Tsu characteristic demonstrates NDC.

\subsection{\label{subsec1_1}Small-signal gain and electric instability in dc biased superlattices}

\begin{figure}
  \includegraphics[scale=0.5]{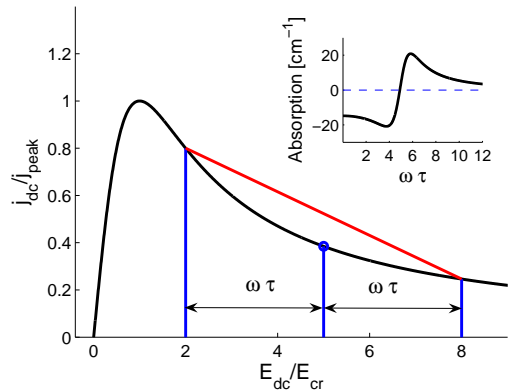}
  \caption{\label{fig:KSSkuva}(Color online) The Esaki-Tsu characteristic.
Figure demonstrates the quantum derivative for the frequency
$\omega=3{\tau}^{-1}$ taken at the operation point
$E_{dc}=5E_{cr}$. Absorption at frequency $\omega$ is proportional
to the difference quotient, Eq.~(\ref{quant-der1}), which is the
slope of the inclined (red online) segment.
Inset: Small-signal gain as a function of $\omega$ for
$E_{dc}=5E_{cr}$. Bloch gain exists almost up to
$\omega\approx\omega_B=5{\tau}^{-1}$ with the maximum of gain at
$\omega\approx\omega_B-\tau^{-1}=4{\tau}^{-1}$. }
\end{figure}
%
We start with the consideration of small-signal gain in  dc-biased
superlattice (\ref{E-mono}). In this case, the analytic
calculation of absorption $\mathcal{A}(\omega)$ gives the
well-known formula of Ktitorov et al. \cite{KSS}. However, having
in mind our future treatment of SL in bi- and polychromatic
fields, it is instructive to introduce here the notion of quantum
derivative and briefly analyze the high-frequency gain using this
tool. The absorption (\ref{absorb-def}) can be represented as the
quantum derivative of the Esaki-Tsu current-field dependence
(\ref{eq:ET}) as \cite{ignatov_q-deriv,wacker_q-deriv,wackerrew}
\begin{equation}
\label{quant-der1} \mathcal{A}(\omega) = \frac{1}{2}\frac{j_{dc}(e
E_{dc} d+\hbar\omega)-j_{dc}(e E_{dc} d-\hbar\omega)}{2 \hbar
\omega} e E_{\delta} d.
\end{equation}
Figure~\ref{fig:KSSkuva} demonstrates the meaning of quantum
derivative (finite difference) in the case of high-frequency probe
field with $\hbar\omega=3\Gamma$ ($\omega \tau =3$) and the
Wannier-Stark spacing $\hbar \omega_B=5\Gamma$ ($E_{dc}=5E_{cr}$).
Geometrically, the quantum derivative represents the slope of
segment with the length defined by the probe frequency $\omega$.
The ends of the segment belong to the Esaki-Tsu curve and their
locations are determined by the choice of working point. In the
quasistatic limit $\omega\tau\ll 1$, we immediately get from
Eq.~(\ref{quant-der1}) the well-known result that the absorption
is proportional to the dc differential conductivity at the
operation point (\ref{esakidiffcond}). On the other hand, as is
obvious from the figure, in order to provide small-signal gain in
the canonic Bloch oscillator the working point must be in the NDC
portion of the Esaki-Tsu characteristic.
\par
The frequency dependence of absorption, calculated using
Eqs.~\ref{quant-der1} and \ref{eq:ET}, is shown in the inset of
Fig.~\ref{fig:KSSkuva}. It illustrates the dispersive profile of
Bloch gain at high frequencies $\omega\simeq\omega_B$ and also the
existence of negative absorption for $\omega\tau\rightarrow 0$.
The frequency corresponding to the gain resonance can be most
easily estimated for the choice of working point satisfying
$E_{dc}\gg E_{cr}$. As is obvious from the geometric meaning of
Eq.~(\ref{quant-der1}), the maximum of gain is achieved when the
left end of the red segment is located at the peak of VI
characteristic. Then a simple geometrical analysis based on
Fig.~\ref{fig:KSSkuva} immediately shows that the maximum of gain
corresponds to the photon energy $\hbar \omega \approx \hbar
\omega_B-\Gamma$ ($\omega\tau\approx\omega_B\tau-1$). This gain
resonance indicates a dissipative quantum nature of the Bloch gain
profile: Asymmetry in the elementary acts of emission and
absorption is caused by the scattering. However, one should always
remember that the feasibility of high-frequency generation in the
canonic Bloch oscillator is complicated by the NDC-related
electric instability.

\subsection{\label{subsec1_2} Photon-assisted peaks and
large-signal gain}

In the case of large amplitude of the probe field, Eq.~\ref{E-mono},
the high-frequency gain is not anymore necessarily connected to the
presence of static NDC, because the ac field can open up new
transport channels leading to the formation of photon-assisted peaks
in VI characteristic \cite{theory_old,unterrainer}. Here the dc
current can be calculated with the help of photon replicas of the
Esaki-Tsu current \cite{wackerrew}
\begin{equation}
j_{dc}^{\omega} (eE_{dc}d) = \sum_l J_l^2(\beta) j_{dc}(eE_{dc}d+l
\hbar \omega) \label{analsolu1}
\end{equation}
and the formula for absorption takes the form
\begin{equation}
 \mathcal{A} = \frac{1}{2}\sum_l
J_l(\beta)\left[ J_{l+1}(\beta)+J_{l-1}(\beta)\right]
j_{dc}(eE_{dc}d+l \hbar \omega), \label{analsolu2}
\end{equation}
where $J_n(x)$ are the Bessel functions, summation is in infinite
limits, $\beta=eE_{\delta}d/\hbar\omega$ and $j_{dc}$ is given by
Eq.~(\ref{eq:ET}). We used the notation $j_{dc}^\omega$ to
distinguish the dc current modified by the action of ac field from
the unmodified Esaki-Tsu current $j_{dc}$. Using Eq.~\ref{analsolu1}
it is easy to calculate the dc differential conductivity,
$\sigma_{dc}^\omega=\partial j_{dc}^\omega/\partial E_{dc}$,
\begin{equation}
 \sigma_{dc}^\omega =\sum_l J_l^2(\beta) \sigma_{dc}(e E_{dc}d+l
\hbar \omega), \label{diff-cond}
\end{equation}
where $\sigma_{dc}$ is given by the formula (\ref{esakidiffcond}).
\begin{figure}
\includegraphics[scale=0.5]{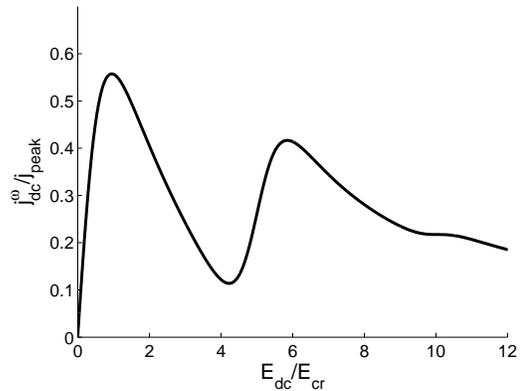}
\caption{\label{fig:Inollavsele} VI characteristic of a superlattice
modified by the strong monochromatic field with $E_{\delta}=5E_{cr}$
and $\omega\tau=5$. Note the appearance of the photon-assisted peak
centered at $E_{dc}/E_{cr}=6$. At this peak the differential
conductivity is positive for
$4=\omega\tau-1<E_{dc}/E_{cr}<\omega\tau+1=6$.}
\end{figure}
%
\par
Figure~\ref{fig:Inollavsele} shows the first photon-assisted peak
in the dependence of $j_{dc}^\omega$ on $E_{dc}$ arising under the
action of strong ac field. Importantly, the left wing of the peak
($4<E_{dc}/E_{cr}<6$) is characterized by PDC. PDC exists if the
Wannier-Stark spacing approximately equals to the photon energy
$|\hbar \omega_B-\hbar \omega|< \Gamma$ (other words, if
$|E_{dc}/E_{cr}-\omega\tau|<1$).
\par
\begin{figure}
  \includegraphics[scale=0.5]{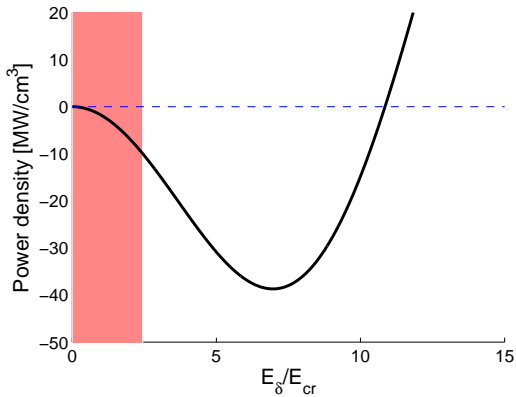}
  \caption{\label{fig:acnfunktiona}
(Color online) Generated power density as a function of the probe
field amplitude $E_\delta$ for fixed $E_{dc}=5.5E_{cr}$ and
$\omega\tau=5$. The dark (red online) segment indicates the range of
negative differential conductivity. If one assumes that the electric
instability due to NDC (red online) at small probe amplitudes does
not prevent the field in the resonator from growing, then the
generation with positive differential conductivity can be achieved
for $E_\delta\in [2.5E_c, 10.8 E_c]$.}
\end{figure}

Now we want to demonstrate the feasibility of large-signal gain in
conditions of static PDC. With this aim we choose the working point
at the left wing of the first photon-assisted peak in
Fig.~\ref{fig:Inollavsele} at $\hbar\omega_B=5.5\Gamma$ ($E_{dc}=5.5
E_{cr}$) and vary the amplitude of the ac field $E_\delta$.
Absorption and, in our notations, power density will be negative if
the SL can generate a high-frequency radiation. The dependence of
the power density on $E_\delta$ is depicted in
Fig.~\ref{fig:acnfunktiona}. In the same figure we also indicated
the range of $E_\delta$ for which NDC is realized at the working
point. We see that the generation definitely can be complicated by
the existence of electric instability at small probe amplitudes:
There always exists a threshold value, which the probe field
amplitude must reach before SL can switch to PDC. However, if the
small-signal space-charge instability can be suppressed by some way,
for example by a very special SL design \cite{savvidis}, the
generated power density at frequency $\omega/2\pi =
5\tau^{-1}/2\pi\approx 4 \ \rm{THz}$  can reach $\sim 40 \
\rm{MW}/\rm{cm}^3$ (Fig.~\ref{fig:acnfunktiona}). For a typical
semiconductor SL it corresponds to the power $\sim 100 \ \mu$W.
%
\begin{figure}
  \includegraphics[scale=0.5]{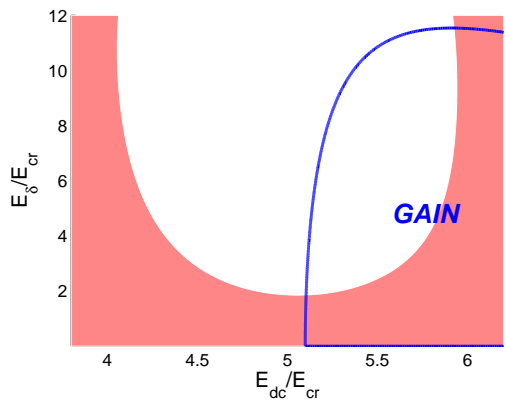}
  \caption{\label{fig:adjustabilities2}
(Color online) Regions of negative differential conductivity (red
online) and gain (marked area bounded by the blue line) for
$\omega\tau=5$. The region of gain overlaps with the region of
\textit{positive} differential conductivity (blank) for $5=\omega
\tau <E_{dc}/E_{cr}< \omega \tau+1=6$ and $E_\delta > 2 E_{cr}$. The
figure indicates the possibility to reach a large-signal
amplification in dc-biased SL without NDC.}
\end{figure}
%
\par
The whole ranges of bias $E_{dc}$ and probe amplitudes
$E_{\delta}$, supporting simultaneously PDC and gain at the first
photon-assisted peak, are presented in
Fig.~\ref{fig:adjustabilities2} for one particular value of the
photon energy $\hbar \omega=5 \Gamma$ ($\omega \tau=5$). This
figure shows that gain without NDC exist if $E_{\delta} \gtrsim 2
E_{cr}$ and $\hbar \omega < \hbar \omega_B < \hbar \omega+\Gamma$
($5<E_{dc}/E_{cr}< 6$).
\par
Similarly, we also calculated the slope of $j_{dc}^\omega(E_{dc})$
dependence and the sign of absorption for many other probe
frequencies. We summarize the requirements to obtain gain without
NDC at the first photon-assisted peak as
\begin{subequations}
\label{allcondA}
\begin{eqnarray}
\hbar\omega &\gtrsim& 2 \Gamma,
 \label{condAa}\\
 E_{\delta} &\gtrsim & 2 E_{cr},
 \label{condAb}\\
 \hbar \omega < & \hbar \omega_B & < \hbar\omega+\Gamma .
  \label{condAc}
\end{eqnarray}
\end{subequations}
The requirements (\ref{condAa}) and (\ref{condAb}) -- which state
that the photon energy must be larger than the scattering induced
broadening and that ac field strength must be no less than two
critical fields -- guarantee the formation of a sharply defined
photon-assisted peak. Note that in the limit opposite to
Eq.~(\ref{condAa}), the photon-assisted peaks never arise (see
\cite{klapandalekseev} and references cited therein). Finally, the
condition (\ref{condAc}) ensures that simultaneously (i) the
absorption is negative $\hbar \omega<\hbar \omega_B$ and (ii) the dc
bias is chosen at the left wing of the photon-assisted peak
corresponding to PDC $|\hbar \omega_B-\hbar \omega|< \Gamma$. For
completeness, we should also mention that PDC and gain can be
simultaneously achieved also by operating close to other
photon-assisted peaks $n \hbar \omega < \hbar \omega_B < n \hbar
\omega+\Gamma$ ($n=2, 3,\ldots$). In this case gain and
modifications in the dc current density occur due to multiphoton
processes, and thus larger amplitudes of the probe field are
required.
\par
We should note that a dynamical response of electrons to THz
radiation can depend on  details of the scattering processes leading
to the relaxation towards a thermal equilibrium. In the case of SLs
with a single miniband, the simple model with two different
scattering constants for the electron velocity $\gamma_v$ and
electron energy $\gamma_\varepsilon$ often becomes more adequate
than the approach based on a single relaxation time
\cite{balance-eqs}. Our numerical calculations indicate that the
inequalities (\ref{allcondA}) well describe the situation for
$\gamma_\varepsilon/\gamma_v\geq 0.5$. However, a large-signal gain
without NDC still can exist even if these scattering constants are
very different (see Appendix~\ref{app:elscatt}).


\section{\label{sec2}Superlattice in bichromatic field}

We turn to the consideration of electron transport in a SL under the
action of a bichromatic ac field
\begin{equation}
\label{E-bichrom} E=E_{dc}+E_{ac}\cos{\Omega
t}+E_\delta\cos{\omega t}.
\end{equation}
We suppose that the ac pump $E_{ac}\cos{\Omega t}$ is strong and the
amplified probe field $E_\delta \cos{\omega t}$ in general can have
an arbitrary amplitude. In the case of generation, the probe field
is a mode of a resonator. We assume that the frequencies $\Omega$
and $\omega$ are incommensurate but both belong to THz frequency
domain ($\Omega\tau\gtrsim 1$, $\omega\tau\gtrsim 1$).
\par
In our earlier discussion, the probe field of a large amplitude
induced some local structures into the VI characteristic of the SL,
which eventually led to the large-signal gain without NDC. The main
idea behind the bichromatic scheme of generation and amplification
is to use an external ac field for the modification of VI
characteristic \cite{theory_old,unterrainer}, thus making it
possible to amplify even a small probe signal without NDC. We will
also show that fields of large amplitude can be amplified in the
same conditions as well.

\subsection{\label{subsec2_1}Main equations}

To begin with we need to calculate the dc current density and
absorption for the bichromatic field (\ref{E-bichrom}) of
arbitrary amplitudes. By using the exact formal solution of
Boltzmann transport equation \cite{chambers} and following earlier
contribution to the formalism \cite{romanov78}, we get
\cite{shorokhov,hyart} for the dc current
\begin{equation}
\label{tasavirta2} j_{dc}^{\Omega\omega}= \sum_{k} J_k^2(\beta)
j_{dc}^\Omega(eE_{dc}d+ k \hbar \omega), \quad \beta=e E_\delta
d/\hbar \omega
\end{equation}
and the absorption of the probe field
\begin{equation}
\mathcal{A}=\frac{1}{2} \sum_k J_k(\beta)\bigg[
J_{k+1}(\beta)+J_{k-1}(\beta)\bigg] j_{dc}^\Omega (eE_{dc}d+k \hbar
\omega). \label{incohabs}
\end{equation}
Here the dc current density modified by the pump field alone is
\begin{equation}
\label{tasavirta3} j_{dc}^\Omega (eE_{dc}d) = \sum_l J_l^2(\alpha)
j_{dc}(eE_{dc}d+l\hbar\Omega),
\end{equation}
where $\alpha=e E_{ac} d/\hbar \Omega$. Next, the dc differential
conductivity $\sigma_{dc}^{\Omega \omega}$ can be easily
calculated from Eq.~(\ref{tasavirta2})
\begin{equation}
 \sigma_{dc}^{\Omega \omega}= \sum_{l,k} J_l^2(\alpha) J_k^2 (\beta)
 \sigma_{dc}(e E_{dc}d+l \hbar \Omega+k
\hbar \omega). \label{diff-cond2}
\end{equation}
NDC and PDC at the operation point $E_{dc}$ correspond to
$\sigma_{dc}^{\Omega \omega}<0$ and $\sigma_{dc}^{\Omega
\omega}>0$, respectively.
\par
In the limit of small probe field ($\beta\ll 1$), we have from
(\ref{tasavirta2}) and (\ref{incohabs}) that $j_{dc}^{\Omega
\omega}\approx j_{dc}^\Omega$ and
\begin{equation}
\label{Aic} \mathcal{A}=\frac{1}{2}
\frac{j_{dc}^{\Omega}(eE_{dc}d+\hbar\omega)-j_{dc}^{\Omega}(eE_{dc}d-\hbar\omega)}{2\hbar\omega}
e E_\delta d,
\end{equation}
where $j_{dc}^\Omega$ is given by Eq.~(\ref{tasavirta3}).
Therefore for a weak probe both dc current and absorption are
determined by VI characteristic modified by the pump field. Since
the formula for small-signal gain in the bichromatic case
(\ref{Aic}) resembles the corresponding formula (\ref{quant-der1})
after the substitution $j_{dc}^{\Omega}\rightarrow j_{dc}$, we can
directly extend the simple geometric analysis of
sec.~\ref{subsec1_1} to the case of more complicated modifications
of VI characteristic caused by the action of strong ac pump.

\subsection{Small-signal analysis: Basic ideas}

\begin{figure}
\centerline{
\includegraphics[scale=0.5]{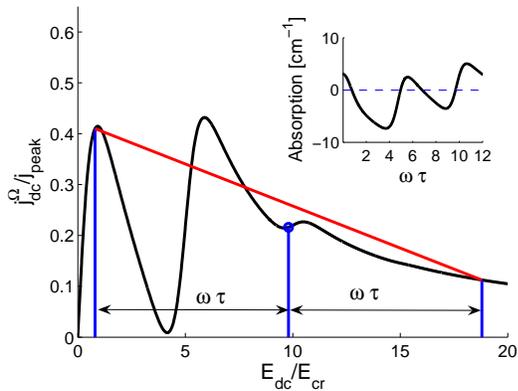}}
\caption{(Color online) VI characteristic of SL under the action
of a strong pump field ($\Omega\tau=5$, $E_{ac}=6 E_{cr}$).
Additionally to the modified Esaki-Tsu peak, two photon-assisted
peaks are visible. The operation point is chosen at the part of
the second peak with a positive slope ($E_{dc}=9.8 E_{cr}$). The
quantum derivative (see Eq.~(\ref{Aic})) at this operation point
is negative for a wide range of $\omega$ values demonstrating the
possibility of small-signal gain without NDC. Inset: Small-signal
absorption as a function of the probe frequency $\omega$ at
$E_{dc}=9.8 E_{cr}$. There are two distinct gain resonances at
frequencies $\omega \tau \approx 3.8$ and $\omega \tau \approx
8.8$. \label{fig:freqshiftex1IE}}
\end{figure}
We want to show how to achieve a high-frequency gain working at
parts of modified VI characteristic with positive slopes. With this
aim we choose ac pump  with large amplitude ($E_{ac}/E_{cr}\gg 1$)
and high frequency ($\Omega\tau\gg 1$). For such pump fields new
local structures in VI characteristic become most pronounced and
simple analytic formulas describing the modifications of dc current
and gain can be derived. Figure~\ref{fig:freqshiftex1IE} illustrates
the dependence of the dc current density on the dc bias which has
been calculated using Eq.~(\ref{tasavirta3}) for $\hbar
\Omega=5\Gamma$ ($\Omega \tau=5$) and $E_{ac}=6 E_{cr}$.
Additionally to the modified Esaki-Tsu peak one can see two new
photon-assisted peaks. The peaks of the current are centered at
\begin{equation}
\label{peaks}
E_{dc}/E_{cr}\approx n \Omega\tau+1
\end{equation}
with $n=0,1,2$. The left wings of the photon-assisted peaks
($4<E_{dc}/E_{cr}<6$ and $9<E_{dc}/E_{cr}< 11$) have positive
slope, i.e. PDC. We found that PDC arises if the Wannier-Stark
spacing approximately equals to the energy of one photon $|\hbar
\omega_B- \hbar \Omega| < \Gamma$
($\Omega\tau-1<E_{dc}/E_{cr}<\Omega\tau+1$) or to the energy of
two photons $|\hbar \omega_B- 2 \hbar \Omega| < \Gamma$ ($2\Omega
\tau-1<E_{dc}/E_{cr}< 2 \Omega \tau+1$). In a more general case
the necessary condition for PDC becomes
\begin{equation}
\label{PDC-bichrom} |\hbar \omega_B-n \hbar \Omega| < \Gamma\quad
(n=0,1,2\ldots).
\end{equation}
The choice of the operation point at these parts of VI should
prevent the development of space-charge instability typical for
systems with NDC. For example, in Fig.~\ref{fig:freqshiftex1IE} we
choose the working point at the left part of second photon-assisted
peak.
\par
Following Eq.~(\ref{Aic}) the calculation of high-frequency
absorption of a weak field requires the finding of quantum
derivative of $j_{dc}^{\Omega}$ at the working point $E_{dc}$. The
geometric meaning of the quantum derivative, which is evident from
Fig.~\ref{fig:freqshiftex1IE}, is similar to the one described in
sec.~\ref{subsec1_1}. Importantly, the slope of (red online)
segment, which is determining the quantum derivative, can be
negative for many $\omega$ providing gain for these frequencies
(Fig.~\ref{fig:freqshiftex1IE}).
\par
It is easy to see that gain has a maximum if the left end of the
segment is located at one of the dc current peaks. Taking into
account that for high frequencies the segment has a small slope
and the peaks are centered at $E_{dc}$ given by Eq.~(\ref{peaks}),
we get that the gain resonances exist for the values of bias and
probe frequency satisfying  $E_{dc}/E_{cr}-\omega \tau \approx k
\Omega\tau+1$ ($k=0,1,2$). In terms of photon energies and
Wannier-Stark spacing, this condition for the gain resonances
takes the form
\begin{equation}
\label{gain_res} \hbar \omega \approx \hbar
\omega_B-k\hbar\Omega-\Gamma.
\end{equation}
The inset of Fig.~\ref{fig:freqshiftex1IE} shows the absorption as
a function of $\omega$ for $\hbar\omega_B=9.8 \Gamma$, $\hbar
\Omega=5\Gamma$ and $E_{ac}=6 E_{cr}$. The gain profile
demonstrates two distinct gain resonances at $\hbar \omega \approx
3.8 \Gamma$ and $\hbar \omega \approx 8.8 \Gamma$, which are well
described by Eq.~(\ref{gain_res}) with $k=1$ and $k=0$.
\par
The dispersive gain profile and the locations of gain maxima
indicate that from a pure quantum perspective, the gain in the
dc-ac-driven SL originates from  scattering-assisted transitions
between quantum mechanical states. In contrast to the Bloch gain
profile in a pure dc-biased SL, the photon sidebands also play their
role here.
\begin{figure}
\centerline{
\includegraphics[scale=0.5]{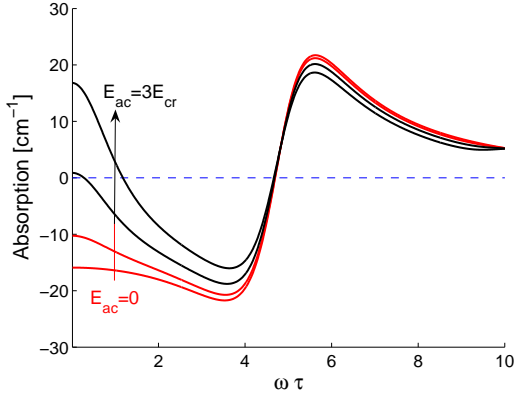}}
\caption{(Color online) Small-signal absorption as a function of the
probe frequency $\omega$ for $\Omega \tau=5$ and different values of
the pump amplitude $E_{ac}/E_{cr}=:0,1.2, 2.1, 3$. The gain profile
near the maximum of gain changes only slightly with an increase of
the pump amplitude. However, the absorption at low frequencies
changes from negative to positive indicating the transition from NDC
(online red curves) to PDC (online black curves) at the operation
point $E_{dc}=4.8 E_{cr}$.} \label{spabsvsx2}
\end{figure}
\par
An interesting question is how the dispersive shape of gain profile
in dc-ac-driven SL is sensitive to the amplitude and frequency of
the pump field. Within the geometrical representation of the quantum
derivative a crossover from gain to absorption corresponds to the
movement of the end of the virtual segment from peak to dip in VI
characteristic. Therefore one can expect that the dispersive shape
of the absorption profile near this crossover is universal as long
as well pronounced local structures in VI do exist. Our numerical
calculations indeed demonstrate that in the vicinity of the
crossover point, which also includes the gain resonance, the gain
profile remains almost unaltered.
\par
Figure~\ref{spabsvsx2} shows the small-signal absorption as a
function of $\omega$ for different amplitudes of the pump, $E_{ac}
\in [0, 3 E_{cr}]$, with the high frequency $\hbar \Omega =5
\Gamma$. We choose the operation point $E_{dc}=4.8 E_{cr}$. If the
ac pump is strong enough,  this operation point belongs to the first
photon-assisted peak with PDC (see Fig.~\ref{fig:freqshiftex1IE},
where $E_{ac}=6 E_{cr}$). For all amplitudes the dispersive Bloch
gain profile at high-frequencies is maintained together with with
the gain resonance at the photon energy $\hbar \omega \approx \hbar
\omega_B-\Gamma=3.8 \Gamma$. However, with increasing pump amplitude
the low-frequency absorption changes from negative to positive; the
switch from NDC to PDC takes place at $E_{ac} \approx 2 E_{cr}$.
Although the electric stability is achieved by a drastic change in
the low-frequency absorption, there is practically no expense in the
high-frequency gain.
\par
To have an analytic formula describing the requirements for the
existence of Bloch gain at PDC in dc-ac-driven SL, we directly
combine the condition for PDC (\ref{PDC-bichrom}) with the formula
for gain resonances (\ref{gain_res}), and get an inequality $(n-k)
\hbar \Omega-2\Gamma <\hbar \omega =
 \hbar \omega_B-k\hbar\Omega-\Gamma< (n-k) \hbar \Omega$.
It can be rewritten as
 \begin{equation}
 \label{gain_res_and_PDC}
 |\hbar\omega-m\hbar\Omega+\Gamma| < \Gamma,
\end{equation}
where the index $m$ ($m=1,2,3\ldots$) marks the $m$-th range of
gain in the absorption profile. We observed numerically that this
naive estimate works surprisingly well for high-frequency pump
fields with amplitudes $E_{ac}\gtrsim 2E_{cr}$. Here the
underlying physics  is the robustness of gain profile at high
frequencies with its simultaneous sensitivity to variations of the
pump field at low frequencies.

\subsection{General conditions for gain: signals of arbitrary
amplitude and positive dc conductivity}

We want to show how to extend our analysis to the case of probe
field which is not weak anymore. Our computations demonstrate that
for parameters satisfying the small-signal gain resonance, the
large-signal gain as a rule also exists. Moreover, we found that the
gain profile near gain resonances stays almost unaltered as the
probe amplitude changes. Furthermore, the important condition
(\ref{gain_res_and_PDC}) works also at large amplitudes of the probe
field. However, simultaneously the analog of Eq.~(\ref{PDC-bichrom})
for the probe field must be also introduced: Wannier-Stark spacing
is approximately equal to an integer multiple of the photon energies
at the probe frequency, $|\hbar \omega_B-l \hbar \omega|<\Gamma$.
\par
Now we can summarize our findings and list all necessary
requirements to be satisfied in order to use the bichromatic
scheme for a generation of high-frequency field under conditions
of electric stability.
\begin{subequations}
\label{allcondB}
\begin{eqnarray}
\hbar\Omega &\gtrsim& 2\Gamma,
 \label{condBa}\\
 E_{ac} &\gtrsim & 2E_{cr},
 \label{condBb}\\
n\hbar\Omega -\Gamma< &\hbar\omega_B& < n\hbar\Omega+\Gamma,
  \label{condBc}\\
\hbar \omega &\gtrsim & 2 \Gamma,
 \label{condBd}\\
l\hbar\omega -\Gamma< &\hbar\omega_B& <l\hbar\omega+\Gamma,
\label{condBe}\\
m\hbar\Omega -2\Gamma< &\hbar\omega& <m\hbar\Omega, \label{condBf}
\end{eqnarray}
\end{subequations}
where in general integers $n$, $l$, and $m$ are different. Some of
these inequalities, like for example (\ref{condBb}) and
(\ref{condBc}), are of course same as derived earlier and here are
listed for completeness. We should note these conditions are
asymptotic and work better the larger is the amplitude
(Eq.~\ref{condBb}) and the higher are the frequencies
(Eqs.~\ref{condBa}, \ref{condBd}).
\par
The requirements (\ref{condBa}-\ref{condBc}) are necessary for the
amplification of a weak probe field at an incommensurate frequency
$\omega$ in conditions of PDC. They ensure that the operation point
is located at the part of well-defined $n$-th photon-assisted peak
having a positive slope. In order the probe field still can be
amplified even if its amplitude is not small anymore, the
requirements (\ref{condBd}-\ref{condBf}) should be taken into an
account. Below we will demonstrate how the requirements
(\ref{allcondB}) work in the cases of THz generation at the first
and second photon-assisted peaks. The effect of different relaxation
constants, $\gamma_v\neq\gamma_\varepsilon$, is analyzed in
Appendix~\ref{app:elscatt}.

\subsubsection{\label{subsec2_2}Gain and generated power density at
the first photon-assisted peak}
\begin{figure}
  \includegraphics[scale=0.4]{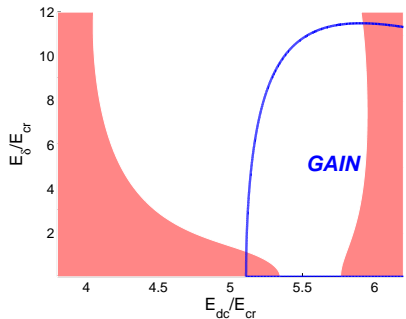}
  \caption{\label{fig:gaindomfig20}
(Color online) Regions of NDC (red online) and gain (marked,
bounded by blue curve) for $E_{ac}=2 E_{cr}$, $\Omega\tau=5.5$,
and $\omega\tau=5$. The figure illustrates the feasibility of gain
in conditions of suppressed electric instability for the wide
ranges of signal amplitudes $E_\delta$ and bias values $E_{dc}$
under the action of  moderate ac pump.}
\end{figure}
\begin{figure}
  \includegraphics[scale=0.4]{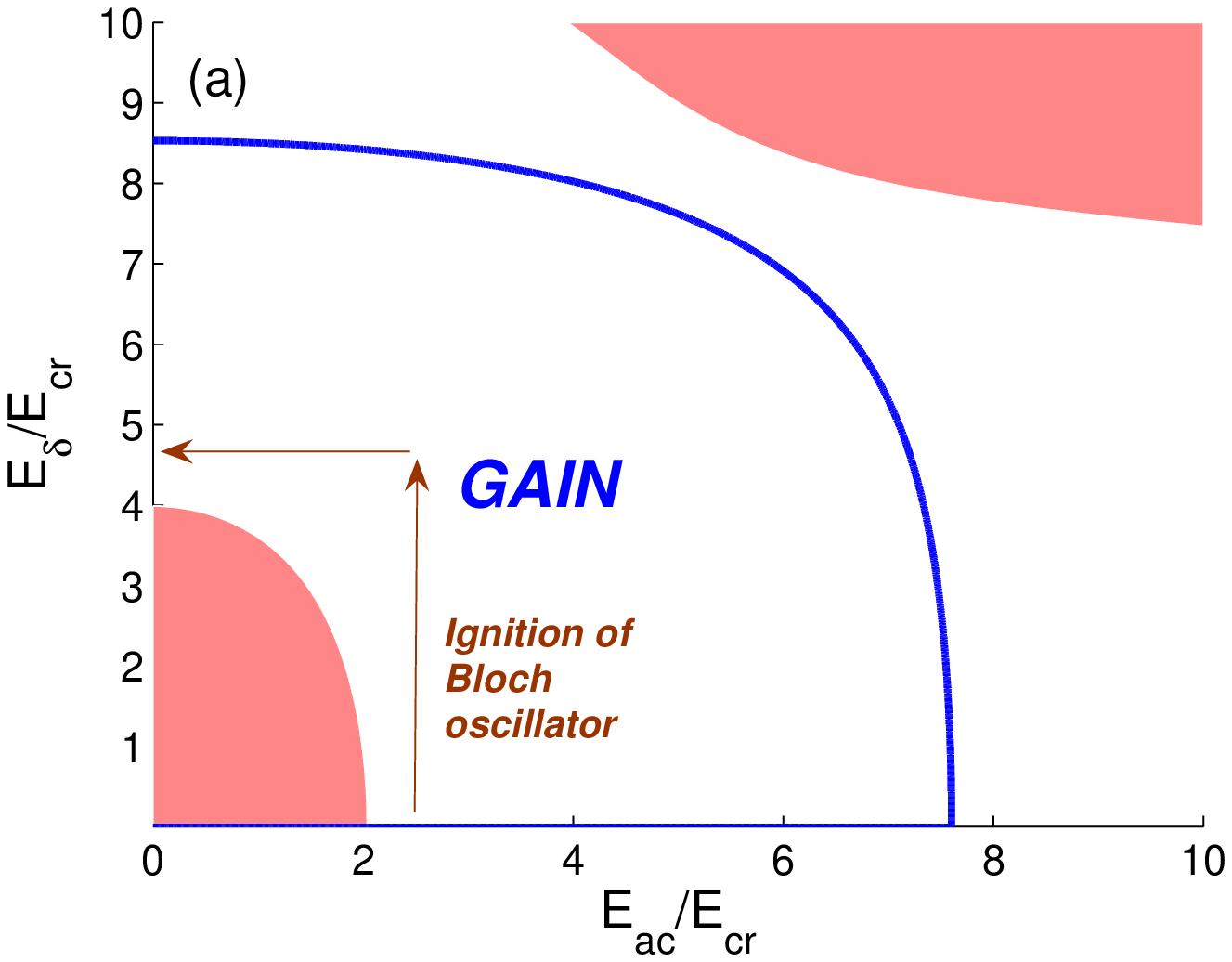}
  \includegraphics[scale=0.4]{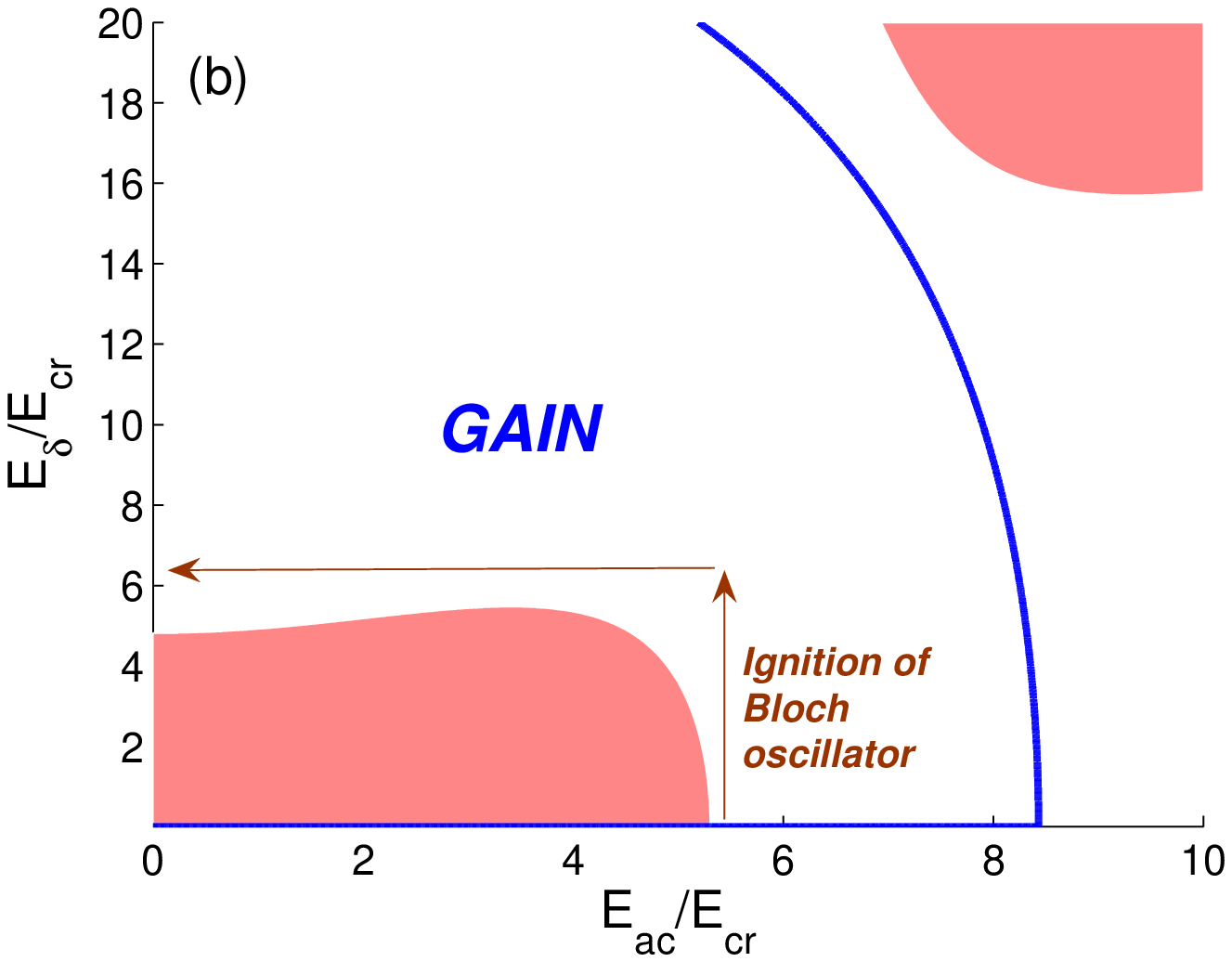}
  \caption{\label{fig:gaindomfigures}
(Color online) Regions of NDC (red online) and gain (marked,
bounded by blue curve) in the plane of pump-probe amplitudes for
(a) $\Omega \tau=5$, $\omega_B \tau=4.8$ and $\omega \tau=4$ and
(b) $\Omega \tau=5$, $\omega_B \tau=9.8$ and $\omega \tau=9$. Gain
can exists without NDC. The arrows schematically illustrate the
use of ac pump in the THz starter for  Bloch oscillator (see
sec.~\ref{Starteri} for details).}
\end{figure}

%
Figure~\ref{fig:gaindomfig20} shows the regions of gain and NDC for
different values of dc bias and probe amplitude $E_\delta$, when the
pump amplitude is fixed (cf. Fig.~\ref{fig:adjustabilities2}). It
demonstrates the possibility to achieve simultaneously PDC and gain
at the first photon-assisted peak [$n=1$ in Eqs.~(\ref{allcondB})]
for the probe frequency in the proximity of the pump frequency
[$m=1$ in Eqs.~(\ref{allcondB})]. In this figure we choose a
relatively low amplitude of the pump $E_{ac}=2 E_{cr}$, which is
close to the threshold amplitude [see Eq.~(\ref{condBb})]. As a
consequence, PDC for $E_\delta\rightarrow 0$ exists for $5.3 E_{cr}<
E_{dc} < 5.8 E_{cr}$, what is a bit narrower than the range $4.5
E_{cr}<E_{dc}<6.5 E_{cr}$ following from Eq.~(\ref{condBc}). On the
other hand, figure~\ref{fig:gaindomfigures}a shows the regions of
gain and NDC in the plane $E_{ac}$-$E_\delta$ for a fixed dc bias.
It demonstrates that gain in conditions of PDC exists for the pump
amplitudes $E_{ac} \in [2E_{cr}, 7.6 E_{cr}]$, when requirements of
Eqs.~(\ref{allcondB}) are satisfied.
\par
\begin{figure}
  \includegraphics[scale=0.5]{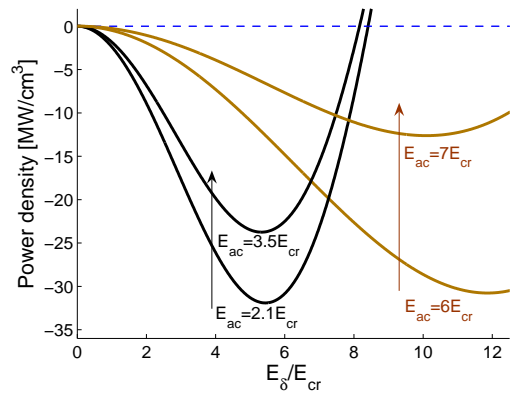}
  \caption{\label{fig:power}
(Color online) Generated power density as a function of the probe
amplitude for several increasing values of the pump. The black
curves correspond to the generation at the first photon-assisted
peak ($\Omega \tau=5$, $E_{dc}=4.8E_{cr}$, $\omega \tau=4$) induced
by the ac pump with $E_{ac}/E_{cr}=:2.1, 3.5$. The brown curves
demonstrate that the generated power at the second photon-assisted
peak ($\Omega \tau=5$, $E_{dc}=9.8E_{cr}$, $\omega \tau=9$) more
strongly depends  on the pump amplitude ($E_{ac}/E_{cr}=:6,7$). As
can be seen from Fig.~\ref{fig:gaindomfigures}, the generation in
both cases can be realized without the electric instability.}
\end{figure}

Figure~\ref{fig:power} shows that the generated power density at
the first photon-assisted peak in the bichromatic scheme slightly
decreases with an increase in the pump amplitude but still remains
comparable with the power density which  can potentially be
generated in the canonic Bloch oscillator (cf.
Fig.~\ref{fig:acnfunktiona}). Moreover, a comparison of
Fig.~\ref{spabsvsx2} and Fig.~\ref{fig:KSSkuva}(inset) allows to
make similar statement concerning the magnitudes of gain in these
two schemes.
%
%

\subsubsection{\label{subsec2_3} Generation with frequency shifting}

By applying a very strong pump field it is possible, due to
multiphoton processes, to exploit other photon-assisted peaks and
reach the generation at frequencies significantly different from
the pump frequency. In the case of second photon-assisted peak
($n=2$ in Eq.~(\ref{condBc})), the frequency of the probe field
can be easily shifted from the pump frequency so that the photon
energy will be close to the gain resonance $\hbar \omega \approx 2
\hbar \Omega - \Gamma$, corresponding to $m=2$ in (\ref{condBf}).
On the other hand, the requirement (\ref{condBe}) can be satisfied
for $l=1$ by tuning dc bias. Therefore, for enough high frequency
of the pump all requirements (\ref{allcondB}) are satisfied, and
amplification of the probe field with both small and large
amplitudes can be achieved in conditions of PDC. Figures
\ref{fig:freqshiftex1IE}, \ref{fig:gaindomfigures}b, and
\ref{fig:power} demonstrate this kind of possibilities for the
pump frequency $\Omega \tau=5$ and dc bias $E_{dc}=9.8 E_{cr}$.


\section{THz starter for Bloch oscillator \label{Starteri}}

It is easy to notice that the large-signal gain in conditions of
PDC is realized at the same parameters of probe fields in both the
monochromatic and bichromatic schemes of amplification. Good
illustration of this comes from the comparison of figures
\ref{fig:adjustabilities2} and \ref{fig:gaindomfig20}. They show
the regions of gain and PDC in these two different schemes for the
same photon energy of the probe field $\hbar \omega=5\Gamma$. In
both cases the regions of gain and PDC overlap approximately for
dc bias values $5.1 E_{cr}<E_{dc}<5.9E_{cr}$ and probe amplitudes
$1.9E_{cr}<E_\delta<11.5 E_{cr}$. However, in the bichromatic case
there also exists an additional overlap region at  small probe
amplitudes for $5.3 E_{cr}<E_{dc}<5.8E_{cr}$ due to the formation
of photon-assisted peak induced by the action of ac pump.
\par
In more general case, we can compare the requirements
(\ref{allcondA}) and (\ref{allcondB}). Assuming that the probe
frequency and dc bias satisfy conditions (\ref{condAa}) and
(\ref{condAc}), we immediately see that the conditions
(\ref{condBd}) and (\ref{condBe} with $l=1$) are also satisfied.
Moreover, the rest of requirements (\ref{allcondB}) can be
satisfied with a proper choice of the pump field.
\par
Therefore it is potentially possible to combine both these schemes
in a single device. The process of domain-free generation in a
resonator with very high-$Q$ will include two stages: In the first
stage, the ac pump is used to excite the field mode with an
amplitude $E_\delta\gtrsim 2E_{cr}$ (requirement (\ref{condAa})).
In the second stage, the pump field is gradually switched off and
sustained generation is performed at a large amplitude of
$E_\delta$. We underline that here the whole switching off process
takes place in conditions of strong gain and PDC.
Figure~\ref{fig:gaindomfigures} illustrates these two stages of
generation utilizing the first (a) and second (b) photon-assisted
peaks. Arrows show a safe pass around the dangerous NDC region in
the plane $E_{ac}$-$E_\delta$, which corresponds to the ignition
of canonic Bloch oscillator.
\par
Since the pump ac field is needed only to the initial stage in order
to suppress the formation of electric domains, and it is then
switched off, it seems reasonable to use THz pulse instead of CW
field in this THz starter for Bloch oscillator. The required
high-power pulses of $100$ ns - $1$ $\mu$s duration can be
generated, for instance, with the help of free-electron laser or
${\rm CO}_2$-laser.
\par
A quite different  situation arises when few-cycle THz pulses are
in use. The field cannot be considered as a monochromatic anymore.
Therefore now we will analyze the feasibility of gain in the case
of polychromatic ac pump. For simplicity, we consider the case of
trichromatic field
\begin{eqnarray}
E_{\rm pump}^{ac}&=&\frac{E_{ac} \cos \Omega t}{2} +\frac{E_{ac}
\cos \Omega_1 t}{4} \nonumber \\&& +\frac{E_{ac}\cos \Omega_2 t
}{4}. \label{pulseeq}
\end{eqnarray}
The amplified probe field continues to be monochromatic $E_{\delta}
\cos\omega t$. We suppose that the pump $\Omega$, $\Omega_1$,
$\Omega_2$ and probe, $\omega$, frequencies satisfy the generalized
incommensurability condition. Namely, we assume that there does not
exist any nontrivial integers $m_i$ ($i=1,2, 3, 4$) such that $m_1
\Omega_1+m_2\Omega_2+m_3\Omega_3+m_4 \omega = 0$. The equation
(\ref{pulseeq}) can be considered as a crude approximation for the
temporal shape of a real broadband THz pulse, where the Fourier
transform of the pulse has a maximum at $\Omega$ and
$|\Omega_2-\Omega_1|$ is related to the bandwidth of the pulse.
\par
The analysis of previous sections can be directly generalized to
the case of polychromatic pump field (\ref{pulseeq}), and we get
for dc current density
\begin{equation}
\label{tasavirta_poly} j_{dc}^{\Omega\Omega_1\Omega_2\omega}=
\sum_{k}J_k^2(\beta)j_{dc}^{\Omega\Omega_1\Omega_2}(eE_{dc}d+ k
\hbar \omega)
\end{equation}
and for absorption of the probe field
\begin{eqnarray}
\mathcal{A}&=&\frac{1}{2} \sum_{k} J_k(\beta)\left[
J_{k+1}(\beta)\nonumber+J_{k-1}(\beta)\right] \nonumber \\&&
j_{dc}^{\Omega\Omega_1\Omega_2}(eE_{dc}d+ k \hbar \omega).
\nonumber
\\ \label{incohabstri}
\end{eqnarray}
Here $j_{dc}^{\Omega\Omega_1\Omega_2}$ is dc current density in SL
modified by the pump field alone
\begin{eqnarray}
\label{tasavirta4} j_{dc}^{\Omega\Omega_1\Omega_2}&=& \sum_{l,m,n}
J_l^2(\alpha/2) J_m^2(\alpha/4) J_n^2(\alpha/4) \nonumber \\&&
j_{dc}(eE_{dc}d+l \hbar \Omega+m\hbar \Omega_1+n\hbar \Omega_2).
\end{eqnarray}
Equations (\ref{tasavirta_poly}) and (\ref{incohabstri}) are valid
for an arbitrary amplitude of the probe field. Remarkably, in the
limit of weak probe field ($E_\delta\rightarrow 0$), the
absorption $\mathcal{A}$ becomes proportional to the quantum
derivative of the pump-modified dc current (\ref{tasavirta4}).
\par
In the limit $\Omega_i\to\Omega$ ($i=1,2$) the trichromatic pump
field (\ref{pulseeq}) becomes momochromatic and therefore one
would expect a restoration of all results of section~\ref{sec2}.
However that is not the case, and the current density and
absorption, given by the equations (\ref{tasavirta_poly}) and
(\ref{incohabstri}), are not reduced to the corresponding
equations of section~\ref{sec2}. The disagreement arises because
the condition of incommensurability is not satisfied in this
limit.
\par
\begin{figure}
\centerline{
\includegraphics[scale=0.5]{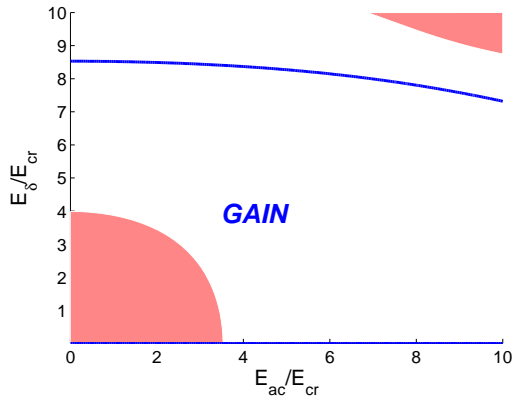}}
\caption{(Color online) Regions of NDC (red online) and gain
(marked) for $\omega_B\tau=4.8$, $\omega\tau=4$ and the trichromatic
pump field (\ref{pulseeq}) with $\Omega\tau=5$, $\Omega_1 \tau=5.5$
and $\Omega_2\tau=4.75$. Comparison of this figure with
Fig.~\ref{fig:gaindomfigures}a demonstrates that the realization of
gain in conditions of PDC under the polychromatic pump requires
larger values of $E_{ac}$ than in the case of momochromatic pump.}
\label{fig:pulse1}
\end{figure}

We computed the regions of NDC and gain for the pump (\ref{pulseeq})
with two different side frequencies $\Omega_1$ and $\Omega_2$, when
the cental frequency $\Omega$ determines the first photon-assisted
peak in VI characteristic. Figure~\ref{fig:pulse1} shows the regions
of gain and PDC in the $E_{ac}$-$E_\delta$ plane. The Wannier-Stark
spacing and photon energies ($\hbar \omega_B=4.8 \Gamma$, $\hbar
\Omega=5\Gamma$ and $\hbar \omega =4 \Gamma$) are the same as used
in figure~\ref{fig:gaindomfigures}a. By comparing figures
\ref{fig:gaindomfigures} and \ref{fig:pulse1}, one observes that the
gain and PDC can be achieved in both cases at similar amplitudes of
the probe field, while a bit larger pump is required in the
trichromatic case.
\par
These results support the possibility to replace a monochromatic
pump with a broadband THz pulse in the starter for Bloch
oscillator. Required intensive T-rays can be produced by a number
of ways, including the use of interdigitated photoconducting
device \cite{T-rays1}, amplifier systems \cite{T-rays2} or from
ultrafast ionizing air \cite{T-rays2} (for recent review, see
\cite{T-rays-rev}).

\section{\label{sec:concl}Conclusion}

We theoretically analyzed the feasibility to reach THz gain in
dc-biased semiconductor superlattice at room temperature using mono-
and polychromatic alternating pump fields. We showed that different
kinds of pump fields can suppress the formation of high-field
electric domains inside the superlattice while still preserve a
broadband THz gain at frequencies incommensurate with the pump
frequencies.
\par
Our approach is based on the utilization of well-known
photon-assisted peaks \cite{theory_old,unterrainer} arising in the
voltage-current characteristics of THz-driven superlattices.
Choice of the operation point at the positive slope of such peak
allows to suppress the electric instability. For the search of
high-frequency gain in conditions of the positive differential
conductivity we employed  simple but powerful geometric
interpretations of the intraminiband absorption formulas.
Combining these analytic tools and numerical simulations, we
demonstrated that the Bloch gain profile in dc-ac-driven
superlattice is robust near the gain resonances.
\par
We suggested to use the robustness property of THz gain in an
effective starter for the canonic (only dc-biased) Bloch oscillator
operating in the electrically stable large-amplitude mode of
generation \cite{Kroemer}. Since THz gain in both the canonic Bloch
oscillator and dc-ac-pumped Bloch oscillator occurs in the same
ranges of dc bias, generation frequency and amplitude, a temporal
application of the ac pump field can be used for domain-free
generation up to the amplitude, which becomes sufficient for a
stable operation of the canonic Bloch oscillator. The pump field,
which ignites the Bloch oscillator, can also be polychromatic. We
predicted that available intensive, broadband THz pulses (T-rays)
can potentially be used in the fast ignition of Bloch oscillator.
\par
We conclude with two remarks. First, in this work we limit our
attention to the case of incommensurate frequencies. If the pump and
probe frequencies are commensurate, an additional coherent term in
the formulas for high-frequency absorption will arise
\cite{Hyart2007}. The sign of the term depends on the value of
relative phase between the pump and amplified fields
\cite{Hyart2007,Hyart2006}. However, at least in the case of
generation with  the monochromatic pump field, the coherent term
always contributes to the high-frequency gain. A more detailed
analysis of this situation, including a consideration of microwave
and sub-THz pump fields \cite{Alekseev2006,romanov-jap06}, will be
presented elsewhere.
\par
Second, we have focused on the local modification of
voltage-current characteristics of superlattices arising when the
Bloch frequency approaches the ac field frequency or its
harmonics. However, similar modifications of the electric
characteristics have been observed in the semiconductor
superlattices subject to the tilted high magnetic field, when the
Bloch frequency is close to the cyclotron frequency or its
harmonics \cite{fromhold}. It is very interesting problem to
consider modifications of the voltage-current characteristics
\cite{bass-magnetic} and analyze the feasibility of THz gain in a
superlattice under the combined action of tilted magnetic and
alternating electric fields.


\begin{acknowledgments}

We thank Ahti Lepp\"{a}nen for cooperation, Andreas Wacker for
attraction out attention to ref.~\onlinecite{wacker_q-deriv} and
Amalia Patane, Alexander Balanov,  Miron Kagan, Alvydas Lisauskas,
Jussi Mattas, Hartmut Roskos, Harald Schneider, Alexey Shorokhov,
Gintaras Valu\v{s}is, Stephan Winnerl for useful discussions. We
are grateful to Feodor Kusmartsev for advices and constant
encouragement of this activity within EU programme. This research
was partially supported by Emil Aaltonen Foundation, Academy of
Finland (101165, 109758), Magnus Ehrnrooth Foundation,
V\"{a}is\"{a}l\"{a} Foundation, and AQDJJ Programme of European
Science Foundation.
\end{acknowledgments}

\appendix

\section{Effect of elastic scattering \label{app:elscatt}}
\begin{figure}
  \includegraphics[scale=0.4]{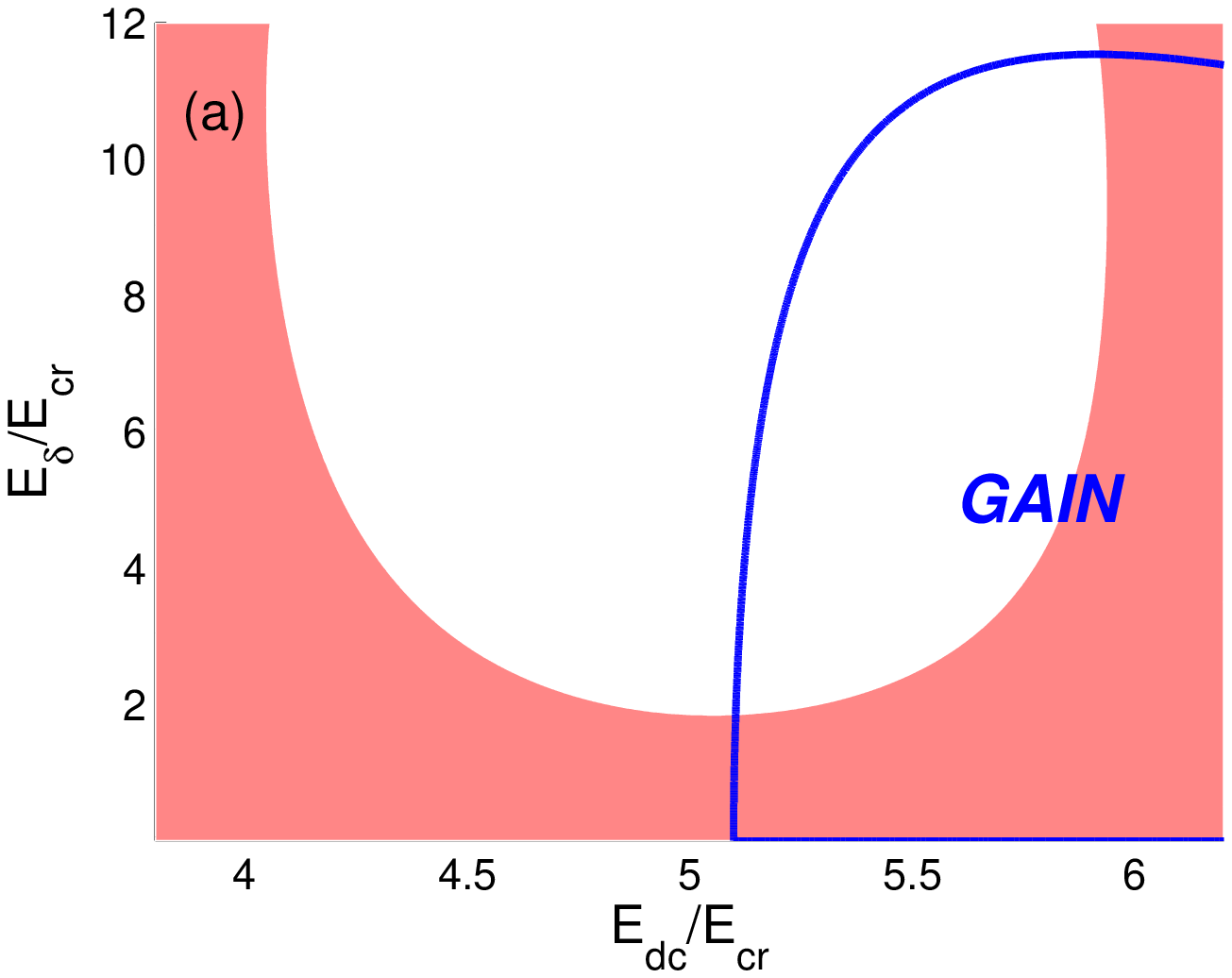}
  \includegraphics[scale=0.4]{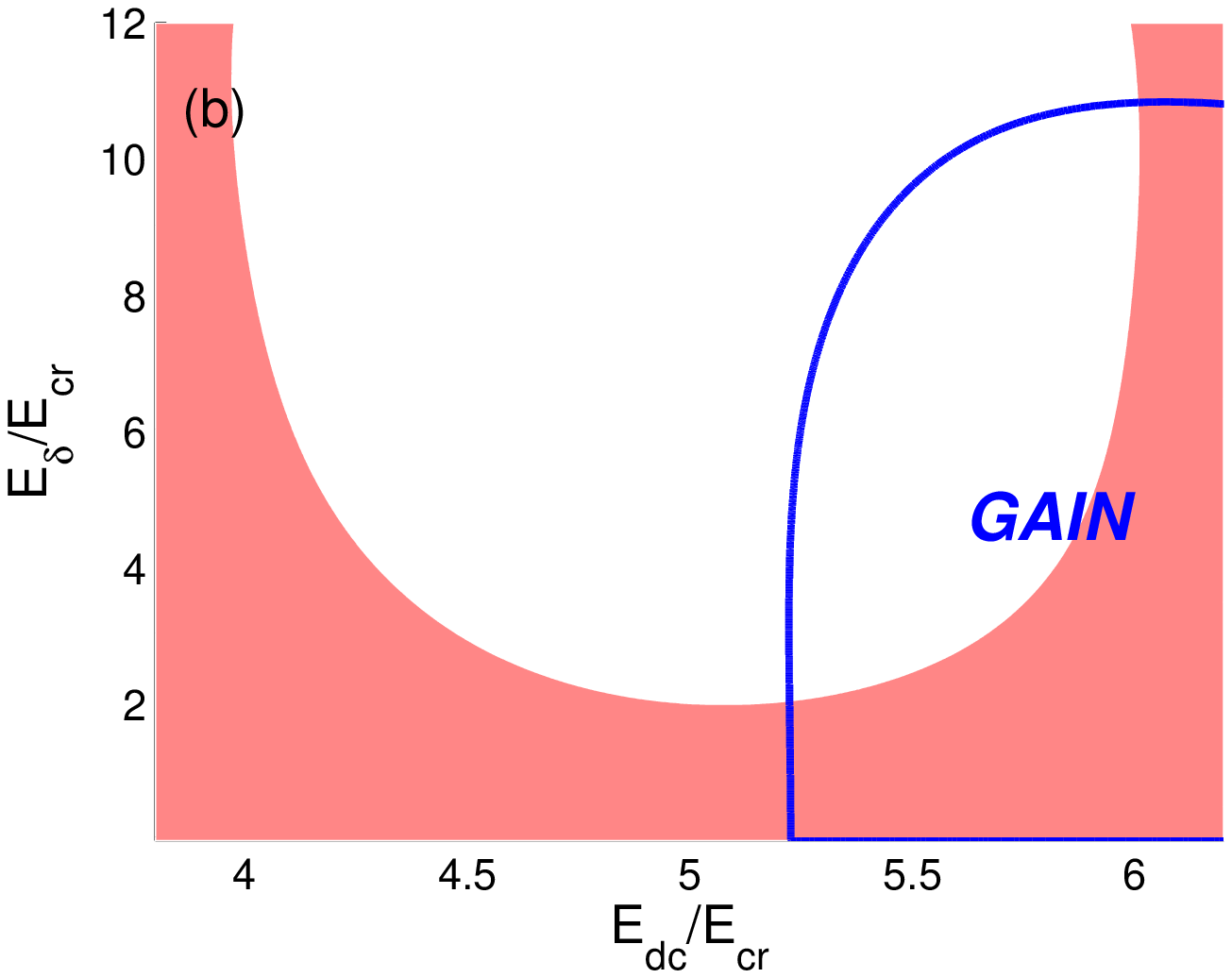}
  \includegraphics[scale=0.4]{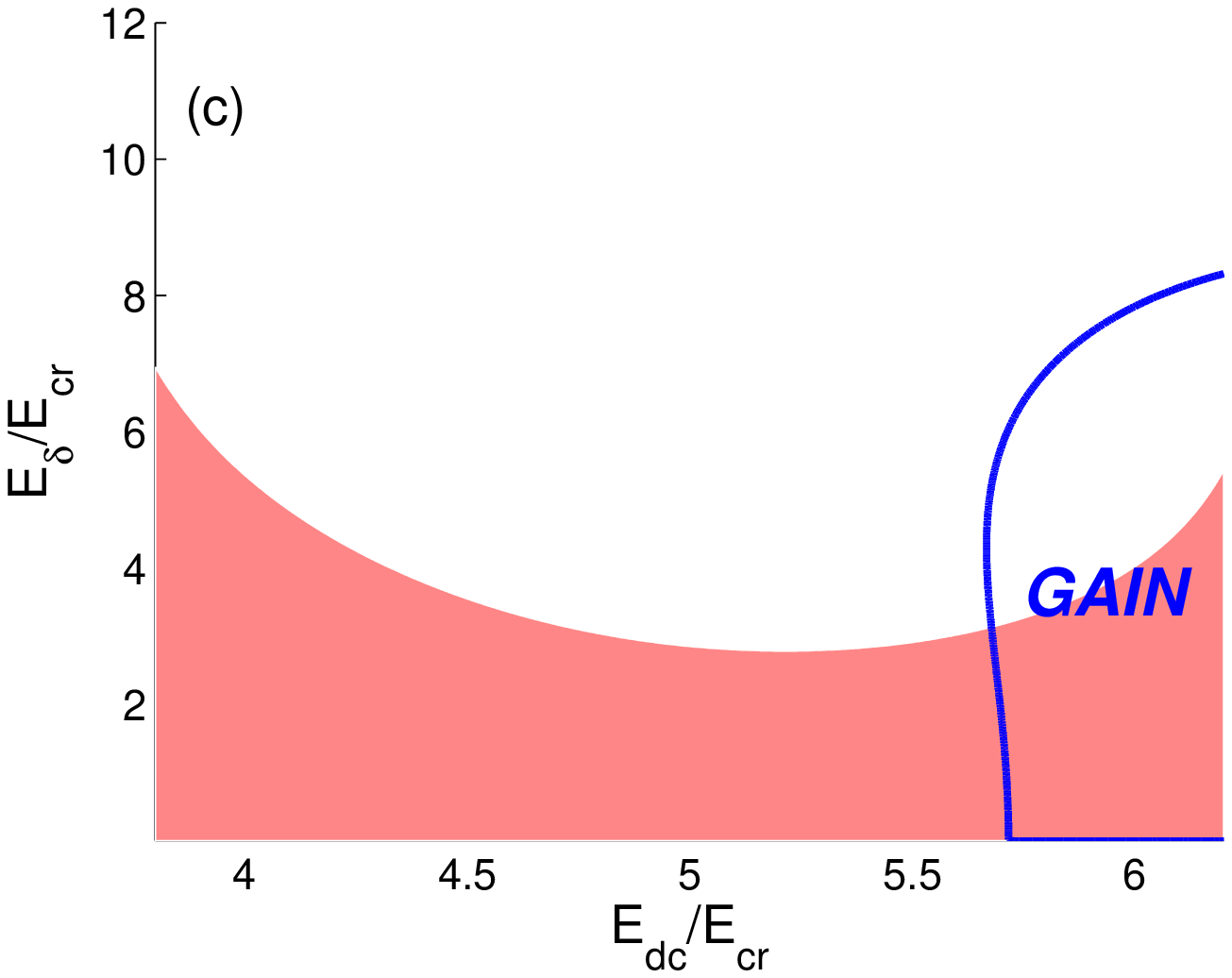}
  \caption{\label{fig:ahtikuva} (Color online) Regions of NDC (red online)
  and gain (marked) in dc-biased superlattice (Eq.~(\ref{E-mono})) for the probe
  field with the frequency $\omega\tau=5$ and for several values of $\nu$: (a) $\nu=1$, (b) $\nu=0.4$ and (c)
$\nu=0.1$. The plots illustrate the feasibility of large-signal
amplification under suppressed electric instability for various
ratios of the scattering constants.}
\end{figure}
\begin{figure}
  \includegraphics[scale=0.4]{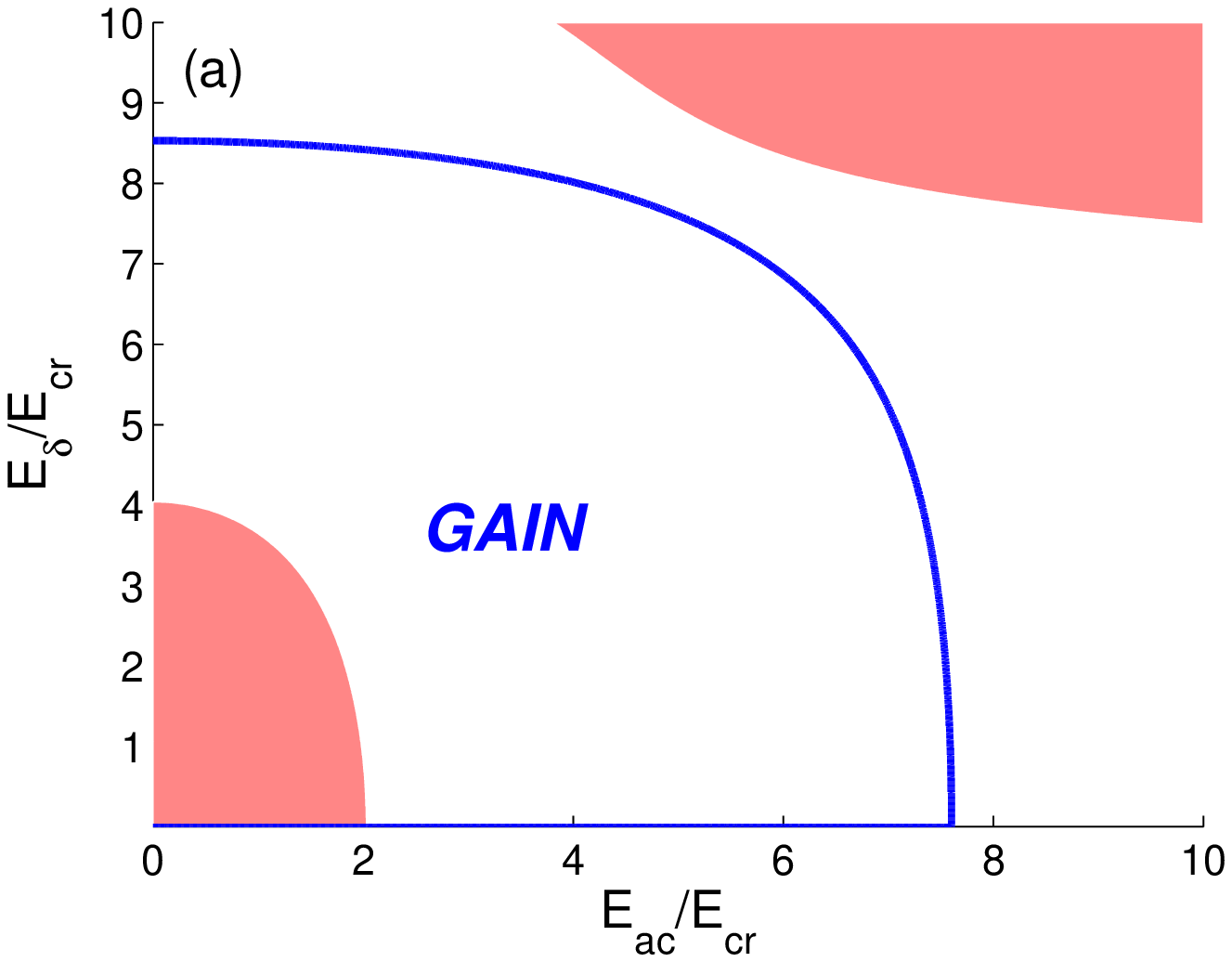}
  \includegraphics[scale=0.4]{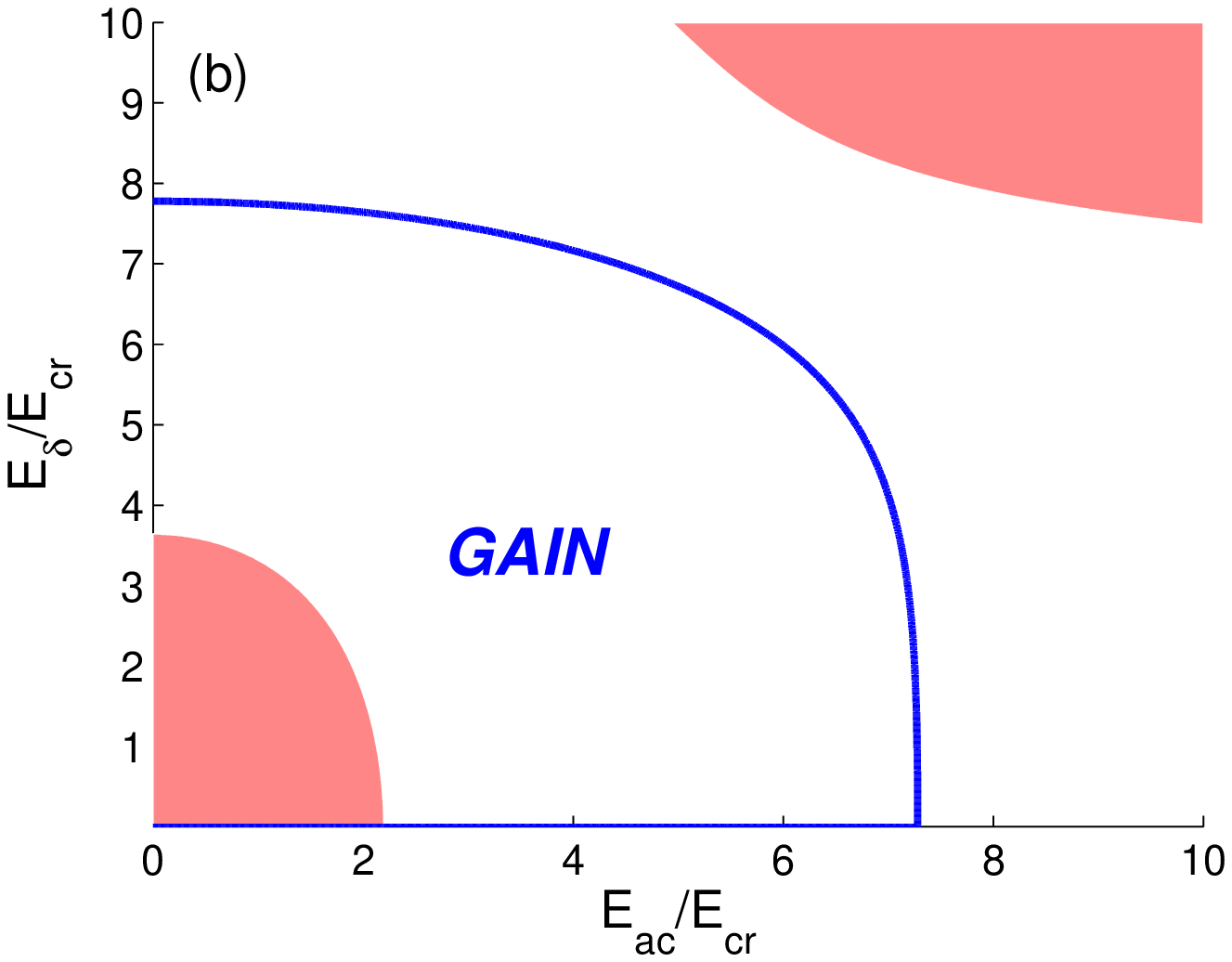}
  \includegraphics[scale=0.4]{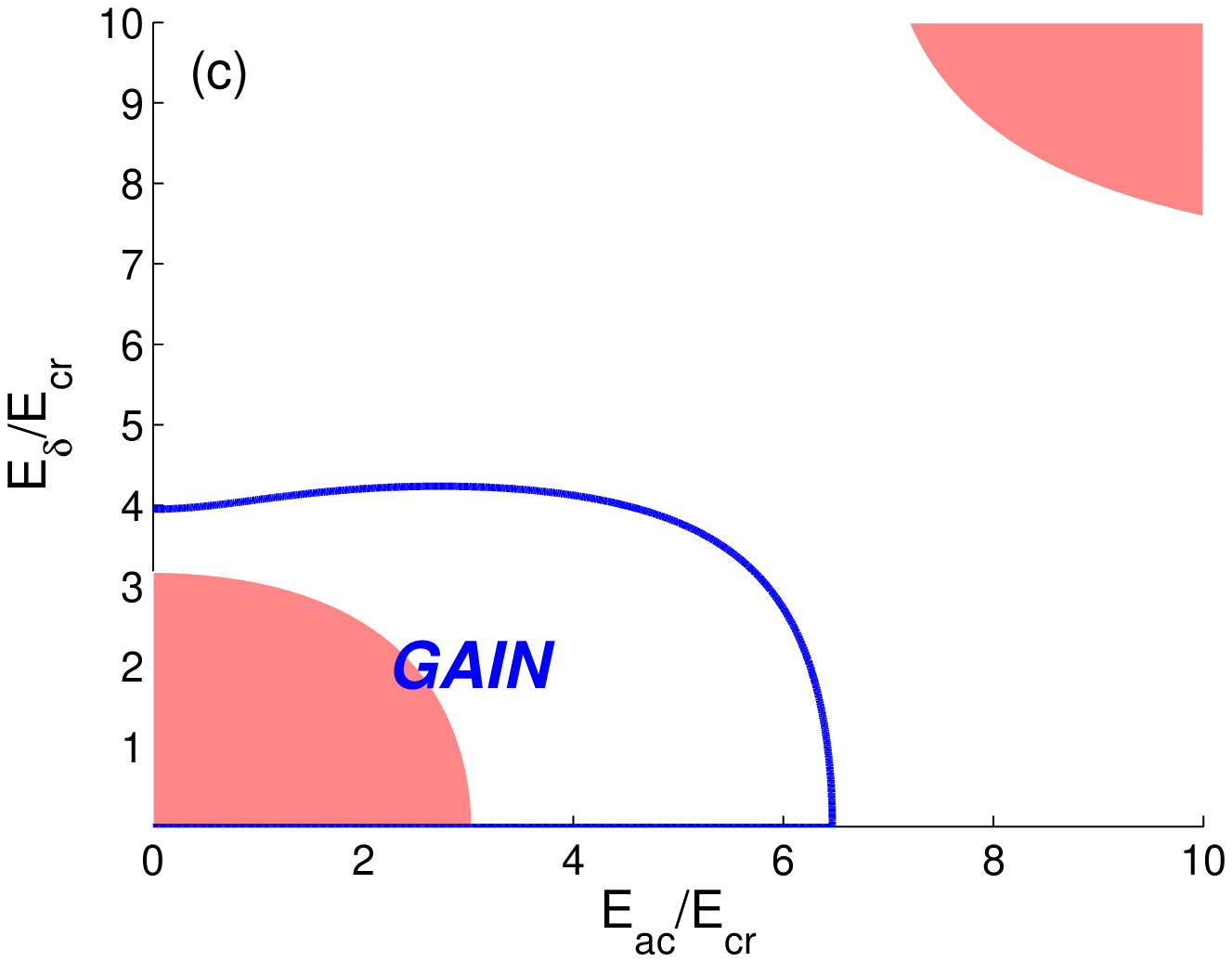}
  \caption{\label{fig:ahtikuva2} (Color online) Regions of NDC (red online)
  and gain (marked) in ac-pumped superlattice (Eq.~(\ref{E-bichrom}))
  for  $\Omega \tau=5$, $\omega \tau=4$,
$\omega_B \tau=4.8$ and for several values of $\nu$: (a) $\nu=1$,
(b) $\nu=0.4$ and (c) $\nu=0.1$. The plots illustrate the
feasibility of generation under suppressed electric instability
for various ratios of the scattering constants.}
\end{figure}
Here we consider the effects of different scattering constants in
both the monochromatic (\ref{E-mono}) and bichromatic
(\ref{E-bichrom}) schemes of THz amplification and generation. The
following SL balance equations can be derived from the Boltzmann
transport equation for a single miniband of SL
\cite{balance-eqs,wackerrew}
\begin{eqnarray}
\frac{d}{dt}j(t)+\frac{d^2 e^2E(t)}{\hbar^2}
\varepsilon(t)&=&-\gamma_v j(t),
\nonumber\\
\frac{d}{dt}\varepsilon(t)- E(t) j(t) &=& -\gamma_\varepsilon
\big[\varepsilon(t)-\varepsilon_{\rm eq}\big], \label{balance}
\end{eqnarray}
where $j(t)$ is the current density and $\varepsilon(t)$ is the
total energy density of electrons within the miniband,
$\gamma_\varepsilon$ and $\gamma_v=\gamma_\varepsilon+\gamma_{el}$
are the phenomenological scattering constants for electron energy
and miniband electron velocity, $\gamma_{el}$ is the scattering
constant describing elastic scattering events. All electrons are
at the bottom of miniband for $\varepsilon=-N \Delta/2$ and the
upper edge of miniband is reached if $\varepsilon=+N \Delta/2$.
The average electron energy in thermal equilibrium
$\varepsilon_{\rm eq}$ depends on the lattice temperature
\cite{wackerrew}.  It is convenient to introduce the mean
scattering time $\tau=1/\sqrt{\gamma_v\gamma_\varepsilon}$ and
ratio of scattering constants $\nu=\gamma_\varepsilon/\gamma_v\leq
1$. According to many experiments \cite{Schomburg} $\nu\gtrsim
0.5$ is a good assumption, although it is not valid for all
superlattices \cite{Hirakawa}.
\par
Solving numerically the SL balance equations (\ref{balance}), we
determine the time-dependence of a steady current for $t\gg\tau$ and
then calculate the dc differential conductivity and absorption of
the probe field. In the case $\nu=1$, we find an excellent agreement
with the earlier analytic results following from the formulas
(\ref{analsolu2}), (\ref{diff-cond}), and (\ref{incohabs}),
(\ref{diff-cond2}). We observe that the photon-assisted peaks occur
in VI characteristics for all values of $\nu\geq 0.1$; however, with
a decrease in the ratio $\nu$ these VI structures become less
pronounced and magnitudes of high-frequency gain are decreasing.
\par
Importantly, the amplification still occurs in approximately the
same ranges of field amplitudes for different values of $\nu$.
Figure~\ref{fig:ahtikuva} represents the ranges of dc bias $E_{dc}$
and ac pump amplitude $E_{ac}$, supporting large-signal gain with
PDC at the first photon-assisted peak induced by the monochromatic
field (\ref{E-mono}). With decreasing $\nu$ the range of amplitudes
for negative absorption shrinks but the range of amplitudes
supporting PDC expands (cf. subplots in Fig.~\ref{fig:ahtikuva}).
The changes are obviously small as long as $\nu \ge 0.4$.
Figure~\ref{fig:ahtikuva2} shows the ranges of pump $E_{ac}$ and
probe $E_\delta$ field amplitudes supporting a generation at the
first photon-assisted peak in the bichromatic scheme. The
suppression of electric domains and the absorption in the
bichromatic field (\ref{E-bichrom}) are based on the same physical
phenomena. Therefore the decrease in $\nu$ affects the bichromatic
scheme in a similar way as in the mochromatic case.

\end{document}